\documentclass{aa}  

\usepackage{lscape}
\usepackage{graphicx}
\usepackage{txfonts}
\begin{document}

   \title{Exoplanets with ELT-METIS I: Estimating the direct imaging exoplanet yield around stars within 6.5 parsecs}

   \author{Rory Bowens\inst{1},
          Michael R. Meyer\inst{1},
          C. Delacroix\inst{2},
          O.~Absil\inst{2}\fnmsep\thanks{F.R.S.-FNRS Research Associate} ,
          R. van Boekel\inst{3},
          S. P. Quanz\inst{4}, 
          M. Shinde\inst{5},
          M. Kenworthy\inst{6},
          B. Carlomagno\inst{2},
          G. Orban de Xivry\inst{2},
          F. Cantalloube\inst{3}, and
          P. Pathak\inst{7}
          }

   \institute{Astronomy Department, University of Michigan, Ann Arbor, MI 48109, USA
         \and
             STAR Institute, Universit\'{e} de Li\`{e}ge, All\'{e}e du Six Ao\^{u}t 19c, 4000 Li\`{e}ge, Belgium
        \and
            Max-Planck-Institut f\"{u}r Astronomie, K\"{o}nigstuhl 17, 69117 Heidelberg, Germany
        \and
            ETH Zurich, Institute for Particle Physics and Astrophysics, Wolfgang-Pauli-Strasse 27, 8093 Zurich, Switzerland
        \and
            Indian Institute of Science Education and Research (IISER) Pune, Dr. Homi Bhabha Rd, Pashan, Pune 411008, India
        \and 
            Leiden Observatory, Leiden University, PO Box 9513, 2300 RA Leiden, The Netherlands 
        \and
            European Southern Observatory, Karl-Schwarzschild-Str. 2, 85748 Garching bei M\"{u}nchen, Germany
             }

   \date{A\&A 2021 in press}

 
  \abstract{Direct imaging is a powerful exoplanet discovery technique that is complementary to other techniques and offers great promise in the era of 30 meter class telescopes. Space-based transit surveys have revolutionized our understanding of the frequency of planets at small orbital radii around Sun-like stars. The next generation of extremely large ground-based telescopes will have the angular resolution and sensitivity to directly image planets with $R < 4R_\oplus$ around the very nearest stars. Here, we predict yields from a direct imaging survey of a volume-limited sample of Sun-like stars with the Mid-Infrared ELT Imager and Spectrograph (METIS) instrument, planned for the 39 m European Southern Observatory (ESO) Extremely Large Telescope (ELT) that is expected to be operational towards the end of the decade. Using Kepler occurrence rates, a sample of stars with spectral types A-K within 6.5 pc, and simulated contrast curves based on an advanced model of what is achievable from coronagraphic imaging with adaptive optics, we estimated the expected yield from METIS using Monte Carlo simulations. We find the METIS expected yield of planets in the N2 band (10.10 - 12.40 $\mu$m) is 1.14 planets, which is greater than comparable observations in the L (3.70 - 3.95 $\mu$m) and M (4.70 - 4.90 $\mu$m) bands. We also determined a 24.6\% chance of detecting at least one Jovian planet in the background limited regime assuming a 1 hour integration. We calculated the yield per star and estimate optimal observing revisit times to increase the yield. We also analyzed a northern hemisphere version of this survey and found there are additional targets worth considering. In conclusion, we present an observing strategy aimed to maximize the possible yield for limited telescope time, resulting in 1.48 expected planets in the N2 band.}


   \titlerunning{Exoplanets with ELT-METIS I}
   \authorrunning{Bowens et al.}
   
   \maketitle
%
\section{Introduction}

The detection and characterization of exoplanets is essential for testing predictive theories of planet formation and evolution. The ultimate goal is to better understand the prospects for life elsewhere in the Universe. Direct imaging is a key technique in the study of exoplanets since it can measure luminosities, constrain temperatures, permit estimates of radii, and investigate atmospheric compositions of exoplanets \citep{traub_oppen_2010}. However, direct imaging is challenging since it requires a high angular resolution given the typical orbital radii of planets as well as the distance from the Sun of typical targets. Since planets located far from their host stars fade as they age (radiating away their heat energy of formation), great sensitivity is also required to detect them. Atmospheric turbulence adds an additional challenge for ground-based observations in achieving the diffraction limit. In addition, high-contrast performance is needed to distinguish the faint light of planets from their bright host stars. 

With recent improvements in adaptive optics (AO), ground-based direct imaging can  closely approach diffraction-limited angular resolution, with improved sensitivity in the background limit. Direct imaging in thermal emission is easier in the contrast limit compared to reflected light for targets around Sun-like stars since the contrast is less stark ($10^{-7}$ in the mid-IR vs $10^{-10}$ in the visible for temperate Earth-sized planets), although reflected light surveys still offer valuable opportunities \citep{refplanets2020A&A...634A..69H}.
In the infrared, the background limit in space is orders of magnitude lower than for ground-based telescopes. However, it is interesting to consider whether larger telescopes on the ground equipped with state-of-the-art adaptive optics can outperform smaller space-based telescopes in the contrast limit. The Near Earths in the Alpha Cen Region (NEAR) experiment recently surpassed estimates of what the James Webb Space Telescope (JWST) will achieve in the contrast limit (\citealt{wagner_2021}, \citealt{beichmann_2020}, \citealt{guyon_2018}) and the NEAR project continues to push sensitivity limits further \citep{pathak2021arXiv210413032P}.
In addition, AO systems are capable of achieving Strehl ratios greater than 80\% from 1-2.5 $\mu$m and over 90\% at wavelengths beyond 3 $\mu$m \citep{davies_kasper_2012}. Indeed, many direct detections of gas giant planets have been achieved in the L band between 3.6 and 4 $\mu$m (\citealt{marois2008Sci...322.1348M}, \citealt{lagrange2009A&A...493L..21L}). The current generation of higher actuator density AO systems is capable of detecting planets in the wavelength range 1 – 2.5 $\mu$m (VLT/SPHERE, Gemini/GPI, and Subaru/SCExAO; \citealt{bez_2008}, \citealt{Macintosh12661}, \citealt{jov_2015}). 

The mid-IR offers unique advantages for characterizing exoplanets compared to shorter wavelengths. Lower temperatures are better probed via longer wavelengths if the observation is sensitive enough to be able to detect such cold objects \citep{heinz_2010}. There have not been many mid-IR AO assisted imagers on the ground. This is in part due to the success of space-based mid-IR platforms in providing unmatched sensitivity. It is also in part due to a perception that AO is not needed to reach the diffraction limit in the mid-infrared. However, the ability to boost contrast is what makes AO indispensable for mid-IR characterization of exoplanets. The mid-IR has already been used with the NEAR project for probing $\alpha$ Cen and other targets on the Very Large Telescope (VLT) but is no longer available (\citealt{pathak2021arXiv210413032P}; \citealt{gayathri_2021arXiv210509773V}). The Nulling Optimized Mid-Infrared Camera (NOMIC) on the Large Binocular Telescope Interferometer (LBTI) is another platform that utilizes AO and nulling interferometry to enhance mid-IR debris disk detection \citep{ertel_2020}. 

Ground-based telescopes in the 30 m class promise sensitivities capable of reaching small ($R < 4R_\oplus$) planets around nearby stars via AO assisted mid-IR direct imaging (\citealt{2013A&A...551A..99Ccrossfield}; \citealt{quanz_2015}). The study by Quanz et al. estimated the exoplanet yield in the 3-10 $\mu$m range for the Mid-infrared ELT Imager and Spectrograph (METIS) on ELT \citep{metis}. By extrapolating preliminary Kepler statistics into the small planet regime and assuming background-limited observations at $2\lambda/D$, they predicted a yield of approximately five exoplanets with one to four Earth radii. Recent advancements in exoplanet demographics and METIS capability predictions have sparked the push for a re-examination. In this work, we use the latest predictions of METIS capabilities alongside updated Kepler statistics to estimate the yield of METIS, particularly in the small planet regime via Monte Carlo simulations. We use a selection of nearby candidate stars to determine the range of contrast and background limited regimes, obtaining a reliable estimate of the yield in exoplanets per star if each band is observed for a 1 hour integration. We then use this estimate to inform an optimal observing plan (including multiple visits) to maximize the yield with METIS on the ELT. We also explore the results of a similar survey conducted in the northern hemisphere. In Section 2, we describe the methods employed to estimate the yield and produce the observation plan. In Section 3, we present the results while in Section 4 we analyze the implications of our study. In Section 5, we draw our conclusions. 

\section{Methods}

To perform our analysis of the METIS yield, we first generated planets as a function of radius and orbital separation from a star. We then simulated high-contrast imaging performance curves for the specific instrument and telescope system.  We selected target stars based on the constraints required for attempting to image small planets in the mid-IR. We assume a single one hour integration per band on each star and ran the simulation 10,000 times. Additionally, we developed methods to predict yields for multiple-epoch observations and background-limited observations of gas giants with residual heat from their formation.

\subsection{Generating synthetic planet populations}
First, we generated synthetic planet populations with known radii and orbits. Occurrence rates based on Kepler data can be used to predict exoplanet populations. We used the NASA ExoPAG SAG13 report\footnote{https://exoplanets.nasa.gov/exep/exopag/sag/}, a meta-study of other Kepler occurrence rate studies that gives the mean number of planets as a function of planet radius and orbital period. We assumed the same planet population irrespective of the host star spectral type.
Our process for generating planets around a star is as follows. We divide the parameter space of period and radius into a grid of cells which we test in a random order. We draw from a Poisson distribution defined by each cell's expectation value to determine if a planet is spawned (see Figure \ref{fig:contrast_with_stars} for the cells with associated expectation value per cell). The planet is assigned a radius and orbital period from a uniform range within the cell. Each cell is tested once per star. If there are already generated planets around that star, we perform a mutual Hill-radii test, as defined in Equation 8 from \citet{Dulz_2020}\footnote{The formula describes stability for circular orbits. We find that after applying the mutual Hill-radii requirement, approximately 2\% of systems generated have at least one occurrence of an overlap of perihelion and aphelion between planets. For this work, even though this assumption is formally inconsistent with our assumed eccentricity distribution, those systems are not removed from the candidate pool.}:
\begin{equation}
\Delta = 2 \left(\frac{a_{P_{out}} - a_{P_{in}}}{a_{P_{out}} + a_{P_{in}}}\right) \left(\frac{3M_{star}}{M_{P_{out}}+M_{P_{in}}}\right)^{1/3}
.\end{equation}
In equation 1, $a_{P_{out}}$ and $a_{P_{in}}$ refer to the outer planet and inner planet semimajor axis values, while the $M$ values refer to the masses of the star, outer planet, and inner planet, respectively.

\citet{SMITH2009381} found five-planet prograde systems with Earth-mass planets were stable on gigayear timescales for $\Delta > 8.5$.
If a newly spawned planet results in a $\Delta < 8.5$ for any of the existing planets, we discard it. We repeat the planet spawning process per star. Solving for the mutual Hill-radii requires the mass of the planets via a mass-radius relation. We follow the work of \citet{ken2017} (in Earth radii and masses)\footnote{Recent work suggests there may be bi-modal M-R relations over a range of radii between super-Earths and gas giants \citep{otegi_2020}.}:
\begin{equation}
\label{small_M}
M = R^{3.57}, R < 1.23
,\end{equation}
\begin{equation}
\label{big_M}
M = 1.48R^{1.69}, R >= 1.23
.\end{equation}
This relationship is derived up to 10 Earth radii. The occurrence rate table extends up to 17 Earth radii but since planets above 10 Earth radii represent only 2\% of planets, we find the best balance of efficiency and accuracy is to assign all of these planets a Jupiter mass. After removing planets that violate the mutual Hill-radii requirement, we find the average system contains 1.72 $\pm$ 1.11 planets.

In addition to a radius and period, we assign planets an eccentricity and a true anomaly. \citet{eylen2019} found, through an analysis of Kepler planets, that eccentricity for systems with single-transits and multi-transits can be described through several types of distributions. We used their half-Gaussian distribution peaked at zero with $\sigma_{multi} = 0.083^{+0.015}_{-0.020}$ for multi-planet systems and $\sigma_{single}$ = 0.32 $\pm$ 0.06 for single-planet systems. We first determine the number of planets in a system and then assign each planet an eccentricity using the appropriate distribution. 
The longitude of the ascending node, the argument of periapsis, and the mean anomaly are all randomly generated from $0$ to $2\pi$. We use the eccentricity and the fraction of the orbital period completed to determine the true anomaly for each planet, following \citet{wright2013}.

The star system as a whole is assigned an inclination with all planets assumed in prograde orbit to that inclination (i.e., the planets revolve around their star in its equatorial plane in the direction of the star's rotation). We tried (when possible) to determine the inclination of the host star systems using estimates of $vsin(i)$ and independently measured rotation periods; for instance, for $\alpha$ Cen A (via applying spectral measurements to astroseismic models resulting in a period of 15.0 $\pm$ 1.0 days \citep{cenA_rot}), $\alpha$ Cen B (via observations of magnetically active spots resulting in a period of 36.2 $\pm$ 1.4 days \citep{cenB_rot}), and Altair (via a fit of a Roche model based on observed properties resulting in an equatorial velocity of 273 $\pm$ 13 km/s \citep{altair_rot_more}).
Inclination ($i$) is then determined via:
\begin{equation}
i = sin^{-1}\left(\frac{vsin(i)}{2 \pi R_{star}/P_{star}}\right)
.\end{equation}
The uncertainty is found via standard propagation of error, assuming variables are independent. $\alpha$ Cen A with a $vsin(i)$ of 2.3 $\pm$ 0.3 km/s has an inclination of 33.9 $\pm$ 7.2 degrees \citep{2005ApJS..159..141V}. $\alpha$ Cen B with a $vsin(i)$ of 0.9 $\pm$ 0.3 km/s has an inclination of 47.0 $\pm$ 22 degrees \citep{2005ApJS..159..141V}. Finally, Altair with a $vsin(i)$ of 203 $\pm$ 3 km/s has an inclination of 48.0 $\pm$ 2.1 degrees \citep{vsini_2005ESASP.560..571G}. For all other systems, the inclination is drawn from a uniform 3D distribution projected into a 2D plane each run, where the probability of an inclination is $p(i) = \sin(i)di$ and the expectation value of an inclination is $\cos(i) = 0.5$.

For the known wide binary star systems, we determine the stable regions for the planets. We use values for the critical semimajor axis coefficients derived by \citet{quarles2020} for a prograde orbit\footnote{This stability criteria was used for the $\alpha$ Cen system however the planets were inserted as discussed above.} based on the original stability limit formula determined by \citet{holman1999}:
\begin{equation}
    a_c/a_b = c_1 + c_2\mu + c_3e_{bin} + c_4\mu e_{bin} + c_5e_{bin}^2 + c_6\mu e_{bin}^2
.\end{equation}
Inserting the stellar mass ratio ($\mu$), the binary semimajor axis ($a_b$), and the eccentricity ($e_{bin}$), we can solve for the critical semimajor axis ($a_c$). Our values for the Alpha Centauri system (Table \ref{bex_data}) are slightly different from those found by Quarles et al., which may be due to differences in the input values. We then use the mass of the target star (Table \ref{nearbylistaa2}) to solve for the critical orbital period. Any planets generated beyond the critical period for a star are discarded. For the purposes of this study, we did not investigate circumbinary planets, as they would be extremely cold and undetectable at $\lambda < 15 \mu m$ for the systems studied here.

\subsection{METIS contrast curves}
Contrast curves for METIS are determined through the High-contrast ELT End-to-end Performance Simulator (HEEPS) \citep{heeps}. 
HEEPS first obtains a temporal series of single conjugate adaptive optics (SCAO) residual phase screens from an end-to-end AO simulation tool. 
HEEPS propagates these SCAO residual phase screens through the METIS high-contrast imaging (HCI) elements (via an optical propagation tool). The radiometric budget from the METIS simulator, the field rotations from the HCI target, and the instantaneous coronagraphic point spread functions (PSFs) are used to produce a mock angular differential imaging (ADI) observing sequence. All targets are assumed to be observed with the Classical Vortex Coronagraph (\citealt{mawet2005}; \citealt{carlomagno2020}). 
Finally, HEEPS uses the Vortex Image Processing package to compute the performance in terms of post-processed contrast \citep{Gomez_Gonzalez_2017}. A median reference PSF is generated and subtracted from all the frames. The frames are derotated and then collapsed along the time axis. The curves are generated with a circularized version of the ELT pupil. Major instrumental effects such as pointing jitter and variable non-common path (NCP) aberrations are included in the computation of the contrast curves, based on the instrumental model built for the METIS Preliminary Design Review in mid-2019. The algorithm used is based on the work of \citet{marois2006}. 

The post-processed 5$\sigma$ contrast ($c$) is defined as the ratio of the noise level ($N$) at a given angular separation to the non-coronagraphic signal of the host star ($S$). The noise level is the standard deviation of the aperture fluxes (using an aperture size $\lambda/D$) measured in as many independent resolution elements as can be defined at the angular separation. The contrast is defined using a t-Student distribution \citep{mawet2014}.
The t-Student distribution results in a significant penalty on the achievable contrast in the small sample limit (i.e., at small $\lambda$/D, where there are few independent resolution elements). This penalty is taken into account in the contrast curves used here. The signal of the host star used is the non-coronagraphic flux. The signal threshold is corrected for the throughput of the post-processing algorithm ($\tau$). The throughput of the post-processing algorithm is determined by injecting fake companions and performing the same analysis to obtain a measurement of the preserved signal in the final image:
\begin{equation}
    5\sigma = c = \frac{5N}{\tau S}
.\end{equation}
Curves are provided with residual errors from known limitations of the SCAO corrections and SCAO plus other instrumental effects such as NCP aberrations and pointing jitter (henceforth "all-effects" curves) for the L, M, and N2 bands. The filters used to generate the curves for each band were the "HCI-L long" filter (3.70 - 3.95 $\mu$m) for the L band, the "CO ref" filter (4.70 - 4.90 $\mu$m) for the M band, and the "N2" filter (10.10 - 12.40 $\mu$m) for the N2 band \citep{carlomagno2020}. The NCP aberrations are errors induced along the path from the AO system to the instrument, such as errors associated with the control of slow drifts, pointing drifts, and pupils drifts. A bottom-up analysis is used to analyze the contributors with a conservative margin applied for each error source. As these are all estimates, it is possible for the true impact of the non-common path errors to be lower than currently predicted. It is also possible some of these effects can be mitigated using enhanced algorithms. In other words, SCAO-only is likely the best that could be hoped for and all-effects is perhaps the worst. All L and M band curves are generated with 0.8 arcsecond fields of view (FOV) radius. All N2 band simulations are calculated for 1.2 arcsecond FOV radius. For each star, an estimated L, M, and N2 mag is used to generate an individual contrast curve set. The L, M, and N2 magnitudes are estimated as discussed in Section 2.3.

\begin{table}[h]\centering
 \caption[]{\label{tab:total_star_list}Candidate stars examined in this work.}
\begin{tabular}{lcccccc}
 \hline \hline
  Star &
  Spec. &
  $m_K$ &
  $m_L$ &
  $m_M$ &
  $m_{N2}$ &

 \\ \hline
        Sirius & A1V$^1$ & -1.35$^2$ & -1.38 & -1.34 & -1.31\\
        Altair & A7V$^1$ & 0.24$^2$ & 0.21 & 0.24 & 0.27\\
        Procyon & F5IV-V$^3$ & -0.65$^2$ & -0.68 & -0.65 & -0.63\\
        $\alpha$ Cen A   & G2V$^4$ & -1.49$^2$ &  -1.52 & -1.47 & -1.48\\
        e Eri & G6V$^6$ & 2.53$^2$ & 2.50 & 2.56 & 2.54 \\
        del Pav & G8IV$^5$ & 2.04$^2$ & 2.01 & 2.07 & 2.05 \\
        $\tau$ Ceti & G8V$^6$ & 1.68$^2$ & 1.65 & 1.71 & 1.69\\
        Omi02 Eri & K0V$^6$ & 2.41$^2$ & 2.38 & 2.44 & 2.42 \\
        70 Oph & K0V$^6$ & 1.79$^2$ & 1.76 & 1.82 & 1.80 \\
        $\alpha$ Cen B   &  K1V$^4$ & -0.60$^2$ & -0.63 & -0.57 & -0.59\\
        36 Oph A & K2V$^4$ & 2.95$^{4*}$ & 2.92 & 2.99 & 2.95 \\
        36 Oph B & K2V$^4$ & 1.84$^{7*}$ & 1.84 & 1.91 & 1.88 \\
        $\epsilon$ Eri & K2V$^6$ & 1.67$^2$ & 1.64 & 1.71 & 1.68 \\
        HD 191408 & K2.5V$^5$ & 3.05$^{8*}$ & 3.01 & 3.09 & 3.05 \\
        70 Oph B & K4V$^9$ & 3.34$^{7*}$ & 3.30 & 3.37 & 3.33 \\
        HD 131977 & K4V$^6$ & 3.10$^2$ & 3.06 & 3.13 & 3.09 \\
        $\epsilon$ Indi & K5V$^4$ & 2.24$^{10}$ & 2.20 & 2.27 & 2.22 \\
        V2215 Oph & K5V$^4$ & 3.47$^{10}$ & 3.42 & 3.50 & 3.45 \\
        \hline
        $\eta$ Cas & F9V$^{11}$ & 1.99$^2$ & 1.96 & 2.00 & 2.00\\
        $\sigma$ Dra & K0V$^6$ & 2.83$^{12}$ & 2.80 & 2.86 & 2.84\\
        HD 219134 & K3V$^6$ & 3.25$^2$ & 3.22 & 3.29 & 3.25\\
        61 Cyg A & K5V$^6$ & 2.68$^{12}$ & 2.64 & 2.71 & 2.67\\
\hline
\end{tabular}
\tablefoot{The list of the 18 ELT candidates with spectral type, apparent K magnitude, and calculated magnitudes. L, M, and N2 magnitudes are calculated from the given K magnitude and spectral type following the work of \citet{pecaut2013}. The top four new candidates for a theoretical northern hemisphere survey at a Hawaii or La Palma site are also given (see Section 2.3 for details) following the dividing line. $\sigma$ Draconis is only visible from La Palma site. Both northern sites can also see Procyon and Altair but only the Hawaii site can see Sirius.\\
\tablefoottext{*}{Converted from a V band magnitude from the given source.}
}
\tablebib{(1) \citet{sirius_spec};
(2) \citet{k-bands}; (3) \citet{procyon_spec}; (4) \citet{spectral_types};
(5) \citet{pav_spec_2006AJ....132..161G}; (6) \citet{list_of_spec}; (7) \citet{ophkband2002A&A...384..180F}; (8) \citet{hd191_vband_2012yCat.1322....0Z}; 
(9) \citet{ophbspec1967AJ.....72.1334C}; (10) \citet{epsilon_k_2003yCat.2246....0C}; (11) \citet{eta_spec}; (12) \citet{k-band_cyg}.
}
\end{table}

The full end-to-end models include photon noise from the star and the background. We assume a one hour exposure time with 40 degrees of rotation for all targets (as if they were all at a declination of -5 degrees, i.e., a zenithal of 19.6 degrees when transiting above the ELT). This is optimistic for some of our targets, which will require slightly longer integrations to achieve the desired rotation. Appropriate sky, instrument, and thermal backgrounds from the radiometric model are included when the curves are generated.

The curves are reported as 5$\sigma$ detection limits versus angular separation. To convert the 5$\sigma$ values and the angular separation of the contrast curves into orbital period ($P$) and limiting planet radius ($R_{L}$) we can detect (the format of the planet occurrence rate tables), we use the following formulas:
\begin{equation}
    P = 2\pi \left(\frac{(d_{star}\theta)^3}{GM_{star}}\right)^{1/2}
,\end{equation}
\begin{equation}
    R_{L} = \left(\frac{5\sigma F_{star} R_{star}^2}{F_{planet}}\right)^{1/2}
.\end{equation}
In the above, $d_{star}$ is the distance to the star, $\theta$ the angular separation of the star and planet, $F_{star}$ the flux of the star at Earth, $F_{planet}$ the flux of the planet at Earth, and $R_{star}$ the radius of the star. We take the telescope diameter ($D_{tel}$) as 37 meters, slightly less than the full edge to edge diameter of 39 meters. A planet's generated period, eccentricity, mass (Eq. \eqref{small_M} and Eq. \eqref{big_M}), and other orbital parameters are used to determine its physical separation from the star, following the deprojecting process detailed in \citet{wright2013}.
We assume the planets are in radiative equilibrium with Bond albedo equal to 0.306, similar to that of Earth and Jupiter. Both star and planet fluxes are derived using the Planck equation for the estimated temperature and the filter response for the specific band. 
We also find the angular separation of the planet and star using the physical separation and the system's inclination. If the planet falls above the contrast curve for the given parameters, it is detected and its values are recorded.

\subsection{Sample selection}
We begin with a volume-limited sample centered on the Sun. Stars must be within 6.5 pc and with declinations between -66 and +16 degrees. The declination range is necessary due to the telescope's latitude and the instrument performance. We balance METIS's zenith angle range (it can perform good SCAO control up to 50 degrees zenith angle) with the assumption that stars will achieve 40 degrees of rotation during an observation. Also taking into account essential high priority targets, we select a minimum allowed altitude angle of 49 degrees. The distance requirement is sensitivity limited to ensure the targets are close enough that there is a reasonable chance a small (< 4 $R_\oplus$) planet might be detected with a one hour integration. The photometric signal-to-noise ratio (S/N) can be written as:
\begin{equation}
    S/N \propto \frac{st_{int}}{\sqrt{st_{int}+n(Bt_{int} + Dt_{int} + \sigma_{RN}^2)}}
\end{equation}
In the above, $s$ is the source signal ($\propto D_{tel}^2 d^{-2}$, where $D_{tel}$ is the telescope diameter and $d$ is the distance to the target), $t_{int}$ the detector integration time, $n$ the number of pixels, $B$ the background signal (whose surface brightness is independent of the telescope diameter in the diffraction limit), $D$ the dark current, and $\sigma_{RN}^2$ the detector read noise. Assuming the background dominates the noise contribution, we can rewrite the equation for a fixed S/N:
\begin{equation}
    S/N_{constant} \propto \frac{D_{tel}^2d^{-2}t_{int}}{\sqrt{t_{int}}}
,\end{equation}
\begin{equation}
    t_{int} \propto \frac{d^4}{D_{tel}^4}
.\end{equation}
Since the integration time goes as the distance to the fourth, it is imperative that we only use nearby stars.
Star selection is limited to K5 stars or earlier with luminosity class IV or V, as these represent stars with a viable chance of planets spatially resolved in thermal emission. Lower luminosity, cooler M dwarfs would have temperate planets at separations that are too small to be resolvable in thermal emission (though are excellent candidates for reflected light observations).

Based on our selection criteria, we found 18 candidate target stars for our survey. We used the known spectral types of the candidates and K band magnitudes to estimate other magnitudes, following the work of \citet{pecaut2013}. They determined conversions to the WISE filters (W1, W2, W3), which correspond approximately to the L, M, and N2, respectively. For stars that lacked a K band mag, we first found the K band mag through a known V band mag. We applied linear interpolation for stars with half step spectral types.

\begin{figure}
    \centering
    \includegraphics[scale=0.55]{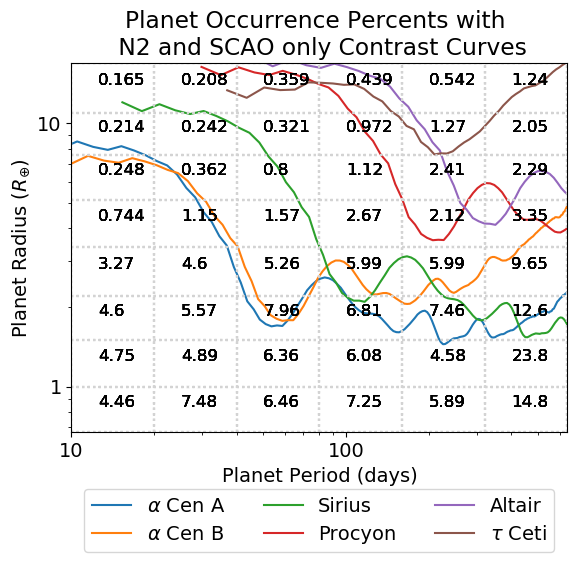}
    \caption{Occurrence rates from the SAG13 report with contrast curves in the N2 band for the top six candidates stars overlaid. The numbers in each cell are the percent chance of a planet with that radius-period combination appearing around a star. Here, $\alpha$ Cen A has the lowest curve and therefore presents the greatest chance of success. There is rapidly decreasing yield from candidates outside of the top six. The curves shown here are for the N2 band with SCAO effects only for a zero inclination, circular system.}
    \label{fig:contrast_with_stars}
\end{figure}

We combine our occurrence rates, contrast curves, and sample star properties and show them in Figure \ref{fig:contrast_with_stars}, where we project contrast curves from Section 2.2 for our targets onto the Kepler occurrence rate grid. We note that the mapping of the contrast onto the Kepler occurrence grid is not intuitive as the 5$\sigma$ contrast is a function of planet radius and planet temperature (which depends on the orbital period). We limit our sample to only the top six candidate stars to improve clarity. We find from preliminary simulations that the top six candidates are responsible for a majority of the detections (over 99\% as shown in Table \ref{tab:obs_plan}).

We also explore appropriate targets for similar observations on a northern hemisphere 30 meter class telescope equipped with a mid-IR camera (e.g., MICHI on the Thirty Meter Telescope (TMT), \citealt{michi}). 
We explore two northern hemisphere sites, assuming identical performance as METIS on the ELT. We adjust our declination range to -21.5 to 60.5 for Hawaii, -12.3 to 69.7 for La Palma. These additional candidates are given in Table \ref{tab:total_star_list}. Although the Alpha Centauri system is out of range at both sites, both can see Procyon and Altair; whereas Hawaii can also see Sirius.\footnote{The GMT on Las Campanas can in principle access all of the top targets for METIS albeit at reduced sensitivity.}

\subsection{Orbital phase space coverage}
We performed additional Monte Carlo simulations to determine the orbital coverage of our survey and explore the value of multiple epoch observations. The orbital phase runs are identical to the standard simulations, but they track which planets were and were not visible during an observation. A time step was performed where the location in orbit for every planet is updated to the new epoch. The analysis was repeated and any newly observable planets were recorded. The code determines an optimal repeat observation time during a set time span (i.e., a time when the greatest increase in yield is expected to occur). It then repeats the process with the initial and repeat observation time to determine the best date for a third observation. This process can be repeated indefinitely, resulting in the theoretical maximum yield for a system. These simulations are performed for a large ensemble of systems that probes the average architectures for the stars, thus identifying the best revisit times given the known parameters of the star.

In this work, we reexamined each star in one month intervals, up to 10 months. Since ideal return times were often within 2 to 4 months, this proved a robust and efficient strategy. The repeat observations still assume a 1 hour integration.

\subsection{Background-limited observations}
We also determined the yield from background limited observations of planets that are not in thermal equilibrium with their stars, that is, where planet luminosity is determined by the heat of formation and model cooling curves as a function of planet mass and system age. These estimates are performed from the limit of the SAG13 data (a period of 640 days) to the edge of the FOV. We begin with the BEX models to estimate the minimum mass of a planet necessary for it to to be robustly detected relative to the background at large orbital separations, assuming the planet's age matches that of the system \citep{bex2019}. We selected (cloud free) atmospheres using the petitCode grid. For the five tested stars, we used the solar metallicity model since -- with one exception (Altair at -0.21 dex, halfway between models) -- they are all within 0.08 dex of solar metallicity (see Table \ref{bex_data}). We provide the  relevant ages in Table \ref{nearbylistaa2}. 

We tested whether internal energy is important for planets closer to their stars (i.e., within a 640 day period). In the most extreme scenario (a 640 day period around Sirius, the youngest target), 50\% of the planets had internal luminosity of the same order as the external luminosity. For the rest of our sample, this value drops to less than 10\%. Therefore, we did not include internal energy for planets spawned using the occurrence rate table (although this may lead to an underestimation of the number of observable planets).

\begin{table*}[h]\centering
 \caption[]{\label{nearbylistaa2}Key parameters for the top six candidate stars selected for this work.}
\begin{tabular}{lcccc}
 \hline \hline
  Star &
  Radius &
  Distance &
  Mass &
  Eff. Temperature \\
  &
  ($R_\odot$) &
  (pc) &
  ($M_\odot$) &
  (K)
 \\ \hline
$\alpha$ Cen A   &  1.225  $\pm$ 0.01$^1$ & 1.3 $\pm$ 0.0$^1$ & 1.173 $\pm$ 0.091$^1$ & 5801 $\pm$ 25$^1$\\
$\alpha$ Cen B & 0.8797 $\pm$ 0.0071$^1$ & 1.3 $\pm$ 0.0$^1$ & 1.021 $\pm$ 0.069$^1$ & 5178 $\pm$ 22$^1$ \\
Sirius &  1.713 $\pm$ 0.009$^2$ & 2.637 $\pm$ 0.011$^{3*}$ & 2.063 $\pm$ 0.023$^4$ & 9845 $\pm$ 64$^{2}$ \\
 Procyon & 1.9190 $\pm$ 0.0150$^1$ & 3.5 $\pm$ 0.0$^1$ & 1.320 $\pm$ 0.100$^1$ & 6543 $\pm$ 25$^1$ \\
 Altair & 1.988 $\pm$ 0.009$^{5}$ & 5.130 $\pm$ 0.015$^{3*}$ & 1.791 $\pm$ 0.018$^{5}$ & 8200 $\pm$ 98$^{5}$\\
 $\tau$ Ceti & 0.8420 $\pm$ 0.0051$^1$ & 3.6 $\pm$ 0.0$^1$ & 0.996 $\pm$ 0.030$^1$ & 5283 $\pm$ 9.4$^1$ \\
 \hline
\end{tabular}
\tablefoot{The stars are ordered from highest to lowest expected yield. These parameters are necessary to calculate if generated planets are above (i.e., visible) or below the METIS contrast curves.\\
\tablefoottext{*}{These values and uncertainties were converted from their original units.}
}

\tablebib{(1) \citet{2005ApJS..159..141V}; (2) \citet{ESASP1200.....E}; (3) \citet{2011PASA...28...58D}; (4) \citet{2017ApJ...840...70B}; (5) \citet{altair2006ApJ...636.1087P}.
}
\end{table*}

\begin{sidewaystable}\centering
 \caption[]{\label{bex_data}Data used to determine background limited yield for the top five candidate stars in this work.}
\begin{tabular}{lccccccccccc}
 \hline \hline
  Star &
  Age &
  Metallicity &
  $\mathcal{M}_L$ &
  $\mathcal{M}_M$ &
  $\mathcal{M}_{N2}$ &
  e &
  a$_{bin}$ &
  $\mu$ &
  Inner Cutoff &
  Outer Cutoff &
  Observable Jovian  \\ 
  &
  (Gyr) &
  (log(Fe/H)) &
  ($M_J$) &
  ($M_J$) &
  ($M_J$) &
  &
  (AU) &
  &
  (AU) &
  (AU) &
  Chance (\%)\\
  \hline
 
$\alpha$ Cen A  & 4.3$^{+1.2}_{-0.3}$  $^1$  & 0.19 $\pm$ 0.017$^1$ & 2.53  & 1.17$^{\dagger}$  & 2.24 & 0.52$^5$ & 23.1$^5$ & 0.47$^5$ & 1.53 & 2.91 & 1.9\\
$\alpha$ Cen B  &  4.3$^{+1.2}_{-0.3}$  $^{1*}$ & 0.15 $\pm$ 0.015$^1$ & 2.53 & 1.17$^{\dagger}$ & 2.24 & 0.52$^5$ & 23.1$^5$ & 0.53$^5$ & 1.46 & 2.58 & 1.6\\
Sirius & 0.242 $\pm$ 0.015$^2$ & -0.071 $\pm$ 0$^{2**}$ & 0.56 & 0.17 & 0.48 & 0.58$^5$ & 19.8$^5$ & 0.33$^5$ & 1.45 & 2.35 & 2.2\\
Procyon & 2.0 $\pm$ 0.0$^1$ & 0.03 $\pm$ 0.017$^1$ & 2.73 & 1.08 & 2.55 & 0.41$^6$ & 15.1$^{6**}$ & 0.31$^6$ & 1.59 & 2.88 & 1.5\\
Altair & 1.096$^{+0.252}_{-0.0}$  $^{3**}$ & -0.217 $\pm$ 0.007$^{4**}$ & 2.08 & 0.93 & 2.08 & - & - & - & 1.77 & 94.8 & 19.0\\
\hline
\end{tabular}
\tablefoot{The distance to the target star and the background limits for METIS for a 1 hour exposure in each band are used to find absolute background limits in each band. Using the solar metallicity BEX model, the minimum observable mass in each band is calculated. For the binary systems, the eccentricity, a$_{bin}$, and mass ratio ($\mu = M_{companion}/(M_{target} + M_{companion})$) are given. The inner cutoff for the target range starts where the occurrence rate table ends (an orbital period of 640 days). The outer cutoff is placed where a binary companion induces instability ($a_{crit}$) or (in the case of Altair) where the FOV limit for METIS occurs. See Figure \ref{fig:space} for a visual diagram of these ranges. The chance of an observable Jovian from the minimum mass (which in all cases was in the M band) over the target range is then reported. See Table \ref{nearbylistaa2} for sources for the distance and mass. We note that in this table, the minimum required masses are given with the symbol $\mathcal{M}$. The total chance of observing at least one Jovian is 24.6\%.\\
\tablefoottext{*}{$\alpha$ Cen B's age is listed as 8$^{+3.6}_{-4}$ in Ref. 1 of this table but assuming they are coeval we adopt the same age as $\alpha$ Cen A.}
\tablefoottext{**}{These values and uncertainties were converted from their original units.}
\tablefoottext{$\dagger$}{These masses required extrapolation of the BEX models but were verified by eye to ensure reliability of the results.}
}

\tablebib{(1) \citet{2005ApJS..159..141V}; (2) \citet{2017ApJ...840...70B};
(3) \citet{1999A&A...348..897L}; (4) \citet{altair2006ApJ...636.1087P}; (5) \citet{binary_info_1989A&A...226..335D}; (6) \citet{procyon_binary_2015ApJ...813..106B}.
}
\end{sidewaystable}

The background limited Vega magnitude for a 1 hour integration is known for each band for METIS: 21.19 for the L band, 18.44 for the M band, and 15.14 for the N2 band. Using distances to each star, an absolute background limit can be determined. These are used to determine the corresponding planet mass that can be seen in the background limited regime for each given system age.

\begin{figure}
    \centering
    \includegraphics[scale=0.85]{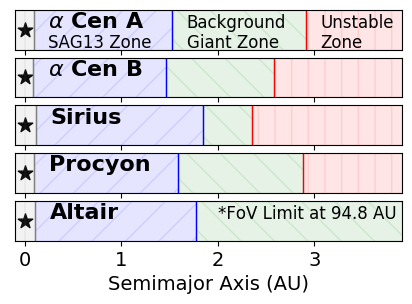}
    \caption{Diagram of the semimajor axis space for the top five stars used in the study. The blue area labeled SAG13 zone represents the period space covered by the SAG13 planets, extending from 10 to 640 days in orbital period. The green area labeled background giant zone is the zone of interest for gas giants which are produced via occurrence rate estimates from \citet{meyer_in_prep}. The red region labeled unstable zone are unstable regions induced by the binary partner (except in the case of Altair); these begin at the critical semimajor axis. The gray area around the star is the space interior of a ten-day period; it is also not used in this study.}
    \label{fig:space}
\end{figure}

For our binary systems, we use the critical orbital period estimated earlier to set a maximum period for generating planets. The minimum period is set equal to the outer edge of our occurrence rate table (640 days). Figure \ref{fig:space} shows the ranges of semimajor axes for each star system that are used in the study. We interpolate the BEX models along magnitude and age in order to determine the minimum mass planet that is observable for the selected band. For $\alpha$ Cen A and $\alpha$ Cen B, minor extrapolation is necessary because of their estimated ages. These extrapolations are verified by eye to ensure they are reasonable. We used the hydrogen burning limit as the upper limit of planet masses (though in no cases were simulated planets that were detected above 20 Jupiter masses.) Finally, using occurrence rate estimates from \citet{meyer_in_prep} which require planet mass range (assuming the planet mass function from \citealt{cumming2008PASP..120..531C} ), orbital separation range, spectral type, and stellar mass, we estimated the percent chance of an observable gas giant planet in the background limit for each system. These rates are determined by combining surveys that utilize all available exoplanet search techniques (cf. \citealt{vigan_2021A&A...651A..72V}).

\section{Results}
\begin{figure}
    \centering
    \includegraphics[scale=0.4]{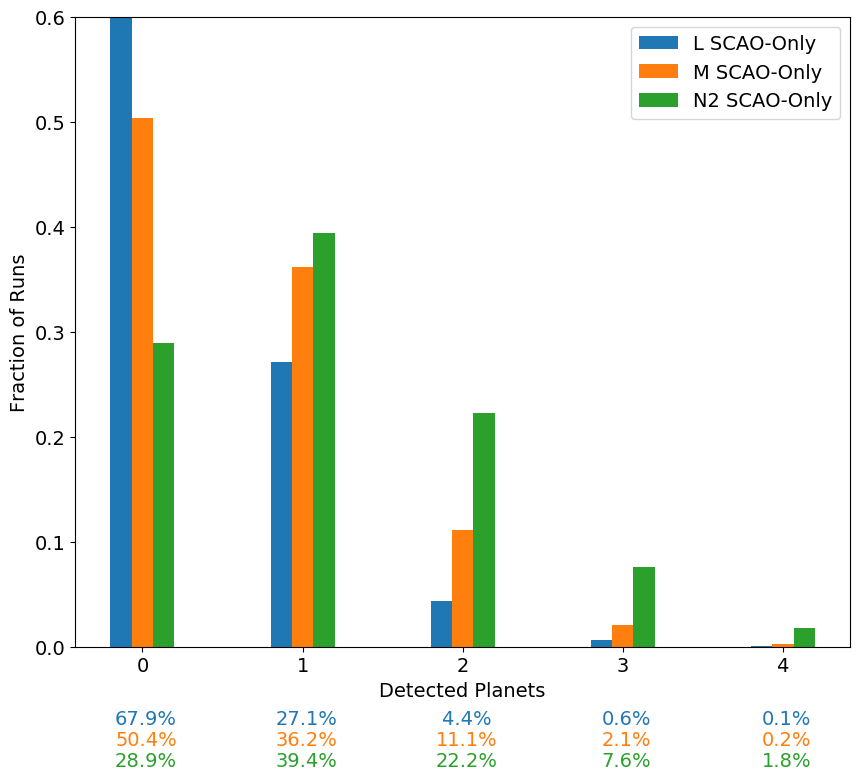}
    \caption{Detected planets around the top five candidate stars for the SCAO-only case with a 1 hour integration per star and only one epoch observed. There is 71.1\% chance of at least one detection in the N2 band.
    }
    \label{fig:scao_hist}
\end{figure}

\begin{figure}
    \centering
    \includegraphics[scale=0.7]{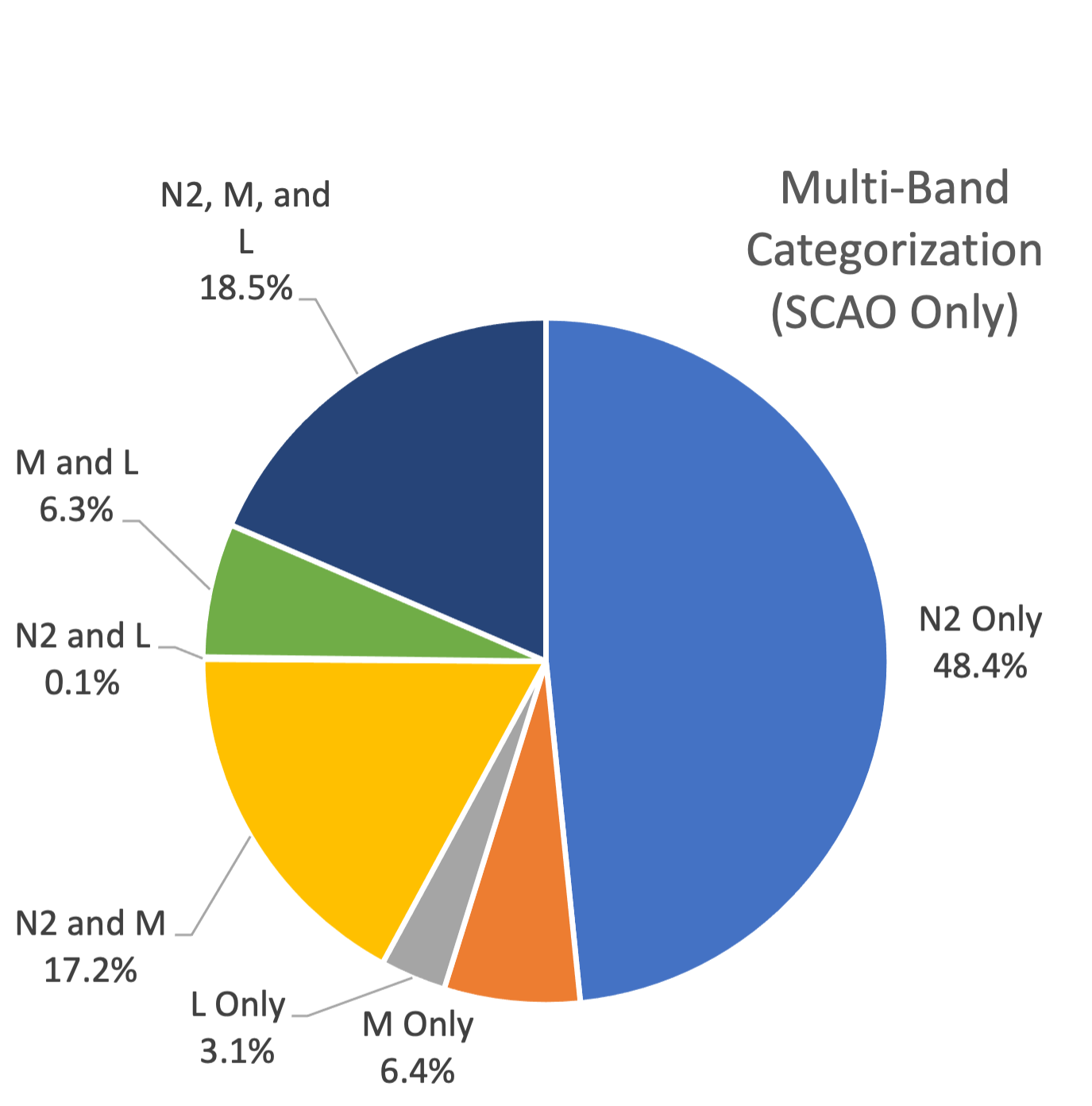}
    \caption{Percentages of detected planets based on observable band(s) for the SCAO-only case with a 1 hour integration per star and only one epoch observed. There is 42.1\% chance of a detection in two or more bands.}
    \label{fig:scao_pie}
\end{figure}

Our results show there is a considerable chance for METIS to yield at least one detection of a small planet (Fig. \ref{fig:scao_hist}) when considering only a single 1 hour observation for each star. The detection will almost certainly occur in the N2 band, although it may occur in M or L as well (Fig. \ref{fig:scao_pie}). In terms of expectation values for SCAO-only (all-effects case in parentheses), the N2 band expected yield is 1.14 (0.49) planets, the M band 0.66 (0.14) planets, and the L band 0.38 (0.06) planets. There is a 71.1\% (40.4\%) chance of detecting one or more planets in the N2 band. Uncertainties are calculated using the bootstrap sampling method. The data pool is sampled for 1000 data points (with replacement) 1000 times and the uncertainties are assumed to be Gaussian. These uncertainties are on the order of 1.5\% for detection chances. However, systematic uncertainties that are due to our assumptions could be higher. There is a 77.3\% (42.9\%) chance of detecting one or more planets using all three bands (an expectation value of 1.41 (0.54) planets) for 1 hour exposures in each band. We also find that 42.1\% (21.2\%) of planets are detectable in two or more bands with uncertainties on the order of 0.1\% using the bootstrap method. Since $\tau$ Ceti only contributed 0.008 (0.002) increase in planet yield in the N2 band, it is used as a cutoff point for the survey, with all results given in terms of the top five stars only.

\begin{figure}
    \centering
    \includegraphics[scale=0.55]{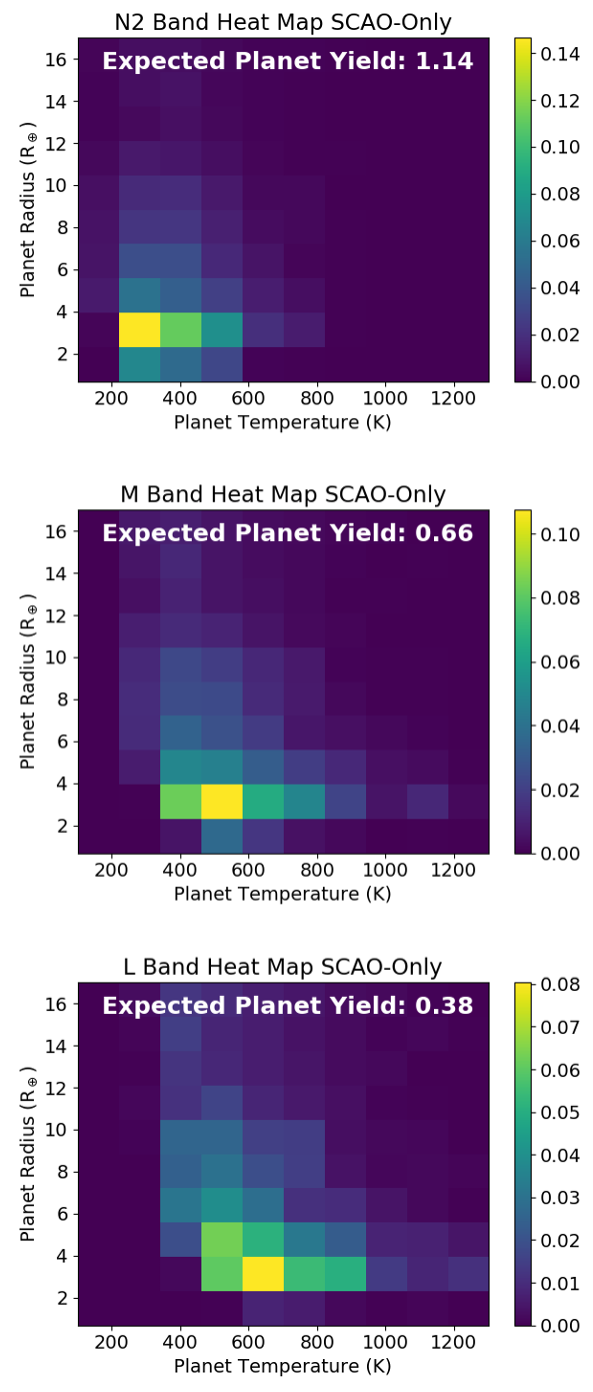}
    \caption{Heat map of detected planet temperature versus radius, separated by band in the SCAO-only case with a 1 hour integration per star and only one epoch observed. The cells give the percentage of planets for the corresponding band found within the temperature-radius range. The expected yield per band over the whole survey is also presented. It can be used in combination with the heat maps to determine the expected yield per cell.
    }
    \label{fig:scao_heat}
\end{figure}

Heat maps are given in Fig. \ref{fig:scao_heat} showing the most likely parameter space for detected planets in terms of radius versus temperature. Planets between 2 to 4 Earth radii with temperatures from 250 to 400 K are the most likely based on the N2 band heat map. This range extends upwards to 500 K. The M band peaks within 2 to 4 Earth radii and extends from 400 K to 700 K. The L band covers a range from 3 to 6 Earth radii with a peak at 3. It also covers a range from 400 K to 1000 K. 84.3\% percent of detections are in N2 for the SCAO case (91.8\% in the all-effects case).

\begin{figure}
    \centering
    \includegraphics[scale=0.5]{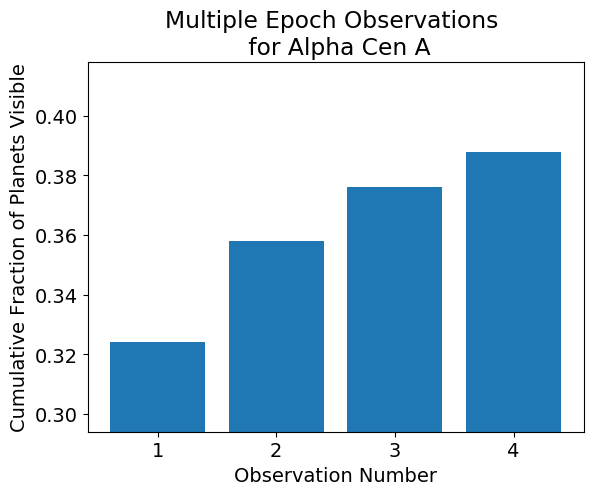}
    \caption{Increase in planet yield for $\alpha$ Centauri A with repeat observations at ideal times for the SCAO-only case. Each bar shows what fraction of the average number of planets generated around $\alpha$ Centauri A have been observed at least once. This process determines the next best observation time from the previous one based on increase in yield (see Table \ref{tab:obs_plan}).
    }
    \label{fig:phase_gain}
\end{figure}

We plot the expected increase in yield from additional epoch observations in Fig. \ref{fig:phase_gain}. Observation dates are optimized for the highest immediate increase in yield from the prior observation. We note that our expected yield from a repeated observation of $\alpha$ Cen A is higher than the expected yield from an initial observation of $\tau$ Ceti; if we stop observing when $\tau$ Ceti is the next target, the total increase in yield from multiple epoch observations raises the N2 SCAO-only case from 1.14 planets (only one of each star) to 1.48 planets (two to four observations per star), as delineated in Table \ref{tab:obs_plan}.

Jovian planet yield per star is shown in the final column of Table \ref{bex_data}. We find the M band has the minimum required mass for the integrations. With the top five candidate stars included, there is a 24.6\% chance to observe at least one Jovian planet in the background limited regime in the M band after a 1 hour integration. Given the long periods (> 2 years), we did not consider multiple epochs.

Candidates for the northern star survey were selected as described in the Section 2 (Methods) for the Hawaii and La Palma locations. A target list of 12 candidates is identified for Hawaii including Sirius, Procyon, Altair, and $\tau$ Ceti. HD 219134, $\eta$ Cassiopeiae, and 61 Cygni A are the three most promising new candidates but their distances and magnitudes make them significantly worse candidates than Altair. For La Palma, a target list of ten candidates is identified (including Procyon and Altair). La Palma can see the three promising candidates from the Hawaii site and a new promising candidate, $\sigma$ Draconis (also worse than Altair). The bottom of Table \ref{tab:total_star_list} gives the estimated magnitudes for these four stars.


\section{Discussion}
There are three substantial caveats to this work and several minor caveats that may impact the results. The validity of occurrence rates used for the close binary systems is a major concern. \citet{Kraus_2016} found that for binary systems with a semimajor axis less than 47 AU, the planet occurrence rates were 0.34 times the given Kepler values which would significantly impact three of the four best target systems. Implementing the decrease for the binaries, we find the expected planet yield for N2 drops from 1.14 to 0.45 assuming contrast curves with errors introduced from the SCAO-only assumption, approximately by one third. Other bands are impacted similarly. Despite the drop, the order of the top five candidates for the study remains the same, although changes would be necessary to prioritize repeated observations (see Table \ref{tab:obs_plan}). However, a more recent study found that although there is reduction in occurrence rate for binaries with a critical radius < 2.5 AU, there is a boosted rate when the critical radius > 3 AU (\citealt{bonavita2020Galax...8...16B}, cf. \citealt{fonta2021arXiv210112667F}). The critical radii determined in our paper are 2.91, 2.58, 2.35, and 2.88 AU for $\alpha$ Cen A, $\alpha$ Cen B, Sirius, and Procyon, respectively. Therefore, the systems may not be impacted as strongly (or at all) at binary separations found for some of our targets. \citet{Quarles_2018} found that the long-term stability for binary systems with two planets is more sensitive to the initial eccentricity state than for a single planet system. To test the impact of such a factor, we perform simulations for the N2 SCAO-only case with the binary systems limited to a maximum of one planet. This decreases the chance of detecting one or more planets for that scenario to approximately 50\%.  Furthermore, the progenitors for Sirius B and Procyon B were closer and larger, indicating a smaller critical semimajor axis. For Sirius \citep{siriusprog1982A&A...114..289D}, we find a critical period of approximately 92 days while for Procyon \citep{procyongprog2015ApJ...813..106B}, we find a critical period of approximately 170 days. The impact of binary evolution is difficult to establish since many factors can influence survival, destruction, or migration \citep{2012ApJ...753...91K}. At the very least the smaller critical periods suggest that these systems may be closer to the one planet limit scenario.

\begin{table}[h]\centering
 \caption[]{\label{tab:obs_plan}Optimized observation plan for the candidate stars in the N2 band.}
\begin{tabular}{lcccc}
 \hline \hline
  Star &
  Observation Number &
  Month &
  Yield Increase
 \\ \hline
    $\alpha$ Cen A & 1 & - & 0.477 \\
    Sirius & 1 & - & 0.277 \\
    $\alpha$ Cen B & 1 & - & 0.263 \\
    Sirius & 2 & 3 & 0.083 \\
    Procyon & 1 & - & 0.061 \\
    $\alpha$ Cen A & 2 & 3 & 0.050 \\
    $\alpha$ Cen B & 2 & 3 & 0.045 \\
    Altair & 1 & - & 0.043 \\
    Sirius & 3 & 6 & 0.038 \\
    $\alpha$ Cen A & 3 & 6 & 0.027 \\
    Procyon & 2 & 2 & 0.022 \\
    $\alpha$ Cen B & 3 & 4 & 0.020 \\
    Sirius & 4 & 11 & 0.018 \\
    $\alpha$ Cen A & 4 & 9 & 0.018 \\
    $\alpha$ Cen B & 4 & 6 & 0.015 \\
    Altair & 2 & 2 & 0.014 \\
    Procyon & 3 & 4 & 0.010 \\
    $\tau$ Ceti & 1 & - & 0.008 \\
    Altair & 3 & 4 & 0.006 \\
    Procyon & 4 & 6 & 0.005 \\
    Altair & 4 & 6 & 0.002 \\

\hline
\end{tabular}
\tablefoot{The list includes the initial observation (month) and three observations at later dates based on a maximum increase in yield each date for the top five stars. Certain stars benefit more from repeat observations, in part because the benefits of repeat observations are inclination dependent. As the chart shows, observing $\tau$ Ceti for the first time provides very low yield. Summing yield prior to $\tau$ Ceti results in an expected yield of 1.48 planets in the N2 band. This observation plan would take approximately 34 hours to perform in N2 only if we assume 50\% efficiency.\\
}
\end{table}
The second substantial caveat is the Kepler occurrence rates used. At the time of SAG13, factors of greater than 2 between different occurrence rates were seen between studies originating from catalog or data product differences (while factors less than 2 could originate from different estimation methods or extrapolation; see the SAG13 report). As seen in Figure \ref{fig:contrast_with_stars}, the highest yield cells mainly consist of those towards the bottom right of the occurrence rate table. If any of these were to change by a sizeable factor, it could significantly impact the final result.  New estimates of occurrence rates based on consistent treatment of completeness and reliability are converging (and consistent with the values from SAG13 for ``Earth--like'' planets), but a full table of updated values is not yet available \citep{Bryson_2020}.  

The third major caveat in this work is the treatment of planets as perfect blackbodies when planet atmospheres, variations in surface albedo, and many other parameters can significantly impact the emergent flux of the planet versus wavelength (\citealt{mendez2017ApJ...837L...1M}, \citealt{makucharticle}). Other caveats include: i) effective temperatures being derived from instantaneous separation from the star, for example 
ignoring the reduction in equilibrium temperature due to eccentricity (which can cause up to a 10\% change \citep{mendez2017ApJ...837L...1M}); ii) assuming that Kepler planets are representative of the solar neighborhood (\citealt{wolfgang2012ApJ...750..148W}, \citealt{tuomi2019arXiv190604644T}); iii) assuming all planets are in prograde orbits with their binary system (see \citet{quarles2020} for details on how this can enlarge or reduce the stability zones); iv) the possible impact of exozodiacal dust emissions \citep{ertel_2020}; v) ignoring the internal energy of the non-background limited planets; and vi) future evolution of the end-to-end contrast curve simulations (for instance as the impact of water vapor and its negative influence on HCI performance are better understood).

In this work, we have found a 71.1\% chance of at least one detection in the N2 band alone with a single hour of integration time per star (assuming SCAO errors only). Using data from the simulations, we produced yield results for each star and epoch, organized from highest to lowest additional yield. The preliminary observation plan should follow the guide in Table \ref{tab:obs_plan} in order to achieve an expected yield of 1.48 planets in the N2 band (if one were to perform all observations until $\tau$ Ceti becomes the next best choice). Similar plans can be constructed for other scenarios (e.g., where we take into account all possible effects that could decrease achievable contrast). Assuming 50\% efficiency, it would take 34 hours to perform the suggested observations in the N2 band, or about 3.5 nights. Accounting for the other two bands, the program would require approximately 11 nights. Integration times can also be increased to improve performance in the background-limited regime  as $t^{-1/2}$. 

Multi-band photometry (and spectroscopy when feasible) enables the estimation of luminosities, equilibrium temperatures, radii, and perhaps atmospheric compositions (including the light element content) for detected planets. Masses may be determined dynamically (e.g., radial velocity or astrometrically) providing bulk density estimates that can be compared with mass-radius relationships derived from radial velocity and transits (e.g. \citealt{otegi_2020}). Furthermore, if an object is also detected in reflected light, we can solve the radius-albedo ambiguity, calculate a global energy budget, and then look for an active greenhouse effect (e.g., \citealt{Sagan52}). Characterization could constrain volatile compositions (e.g., C, N, and O) and the presence or absence of a dense hydrogen+He atmosphere \citep{lammer2020SSRv..216...74L}. Such a discovery would be a major breakthrough in exoplanet characterization and point the way towards robust biosignature searches \citep{kiangdoi:10.1089/ast.2018.1862}. 

A major question in planet formation theory is how the observed exoplanet yields versus orbital period and planet radius compare to theoretical predictions (cf. \citealt{Chambers_2018}, \citealt{alibert2005A&A...434..343A}, \citealt{Ida_2004}). Unfortunately, the small scope of our survey makes it unlikely that these results would have profound implications for existing models. Instead, any results may be more a reflection of the dynamical configuration of specific systems: planet formation zones, migration paths, and stable regions. In any case, a mid-IR camera on an ELT could make other breakthrough exoplanet discoveries such as observing protoplanets forming in a protoplanetary disk. These targets have multiple emission components so mid-IR data would serve as an excellent complement to near-IR and mm-wave studies of these regions \citep{quanz_2015}. A mid-IR camera could also directly detect planets discovered dynamically (and these surveys can be optimized for exposure time and ideal epoch). Examples include the Jupiter-mass planet around $\tau$ Ceti \citep{kervella2019A&A...623A..72K}, an approximately 0.78 Jupiter-mass planet around $\epsilon$ Eridani \citep{mawet2019AJ....157...33M}, and an approximately 3 Jupiter-mass planet around $\epsilon$ Indi A \citep{feng2019MNRAS.490.5002F}. The planet around $\epsilon$ Eridani and the planet around $\epsilon$ Indi A should be easily detected thanks to the young system age and high mass, respectively. Estimating with the magnitudes from the solar metallicity BEX models, the planets for $\epsilon$ Eridani and $\epsilon$ Indi A would only take several minutes to detect given their expected masses and ages. The planet around $\tau$ Ceti would take longer due to its mass and age, and assuming it has the same low metallicity as its host star. Using the BEX -0.4 model since $\tau$ Ceti has a metallicity of -0.36 dex \citep{2005ApJS..159..141V}, we estimate it would require a 4 hour exposure to detect a Jupiter-mass planet given a system age of 7.24$^{+4.78}_{-2.88}$ Gyr \citep{1999A&A...348..897L}. Furthermore, the growing list of dynamically discovered planets from radial velocity legacy surveys and Gaia astrometry will provide several mature, cold planets that can be characterized in mid-IR using METIS (\citealt{blunt2019AJ....158..181B}, \citealt{sozzetti2014MNRAS.437..497S}). Finally, METIS should be capable of imaging many young gas giants at or beyond the runaway accretion phase \citep{ireland_10.1093/mnras/stz2600}.

\section{Summary and conclusions}
In this study, we combine the latest estimates of the capabilities of METIS with Kepler occurrence rates in order to predict the exoplanet direct imaging yield of a future METIS survey of targets within 6.5 pc from super-Earths to gas giants in thermal emission. Using Monte Carlo simulations, we predict the expected yield from METIS for our top five candidate stars in the L, M, and N2 bands. For the SCAO error-only case, we find:

\begin{enumerate}
  \item The N2 band outperforms the M and L bands for our mock survey parameters.
  \item There is a 1.14 expected planet yield in the N2 band for 1 hour exposures with periods < 640 days and radii as small as 1 $R_\oplus$ (1.41 expected planet yield if the N2, M, and L bands each have 1 hour exposures).
  \item There is a 42.1\% chance of observing the same planet in two or more bands.
  \item There is an approximately 24.6\% chance of observing a Jovian planet in the background limit in the M band.
\end{enumerate}

Furthermore, we use multiple epoch testing to produce a preliminary observation plan organized by expected yield. We applied a similar analysis to the two possible TMT sites, assuming similar performance as METIS on the ELT. Finally, our results indicate that there is significant discovery space for mid-IR cameras on the ELTs in imaging planets around the very nearest stars.
\begin{acknowledgements}
This project was made possible through the support of a grant from Templeton World Charity Foundation, Inc. The opinions expressed in this publication are those of the authors and do not necessarily reflect the views of the Templeton World Charity Foundation, Inc.

This project has received funding from the European Research Council (ERC) under the European Union’s Horizon 2020 research and innovation programme (grant agreement No 819155), and from the Wallonia-Brussels Federation (grant for Concerted Research Actions). 

Part of this work has been carried out within the framework of the National Centre of Competence in Research Planets supported by the Swiss National Science Foundation. SPQ acknowledges the financial support of the SNSF.

We are grateful to M. Kasper (ESO),  C. Marois (HIA), C. Packham (UTSA), J. Males (UofA), as well as the FEPS Research Group at UM for helpful conversations.
\end{acknowledgements}

\bibliographystyle{aa}
\bibliography{bibfile}

\end{document}